\documentclass[prl,amsmath,amssymb,aps,superscriptaddress,floatfix,reprint,notitlepage]{revtex4-1}
\usepackage{natbib}
\usepackage{graphicx}
\usepackage{color}
\usepackage{dcolumn}
\usepackage{bm}
\usepackage[utf8]{inputenc}
\usepackage{amssymb}
\usepackage{color,amsmath}
\usepackage{mathtools,ulem}
\usepackage{soul,xcolor} 
\setstcolor{red} 
\usepackage{epstopdf}
\usepackage{scalerel}
\usepackage{hyperref}
\hypersetup{
 pdfnewwindow=true, colorlinks=true,
 linkcolor=blue, anchorcolor=blue,
 citecolor=blue, filecolor=blue,
 menucolor=blue, urlcolor=blue}

\begin{document}

\title{Topological charge density waves at half-integer filling of a moir\'e superlattice 
}
\author{H. Polshyn} 
\affiliation{Department of Physics, University of California, Santa Barbara, CA 93106}
\author{Y. Zhang} 
\affiliation{Department of Physics, University of California, Santa Barbara, CA 93106}
\author{M. A. Kumar}
\affiliation{Department of Physics, University of California, Santa Barbara, CA 93106}
\author{T. Soejima}
\affiliation{Department of Physics, University of California, Berkeley, CA 94720, USA}
\author{P. Ledwith}
\affiliation{Department of Physics, Harvard University, Cambridge, MA~02138, USA}
\author{K. Watanabe} 
\affiliation{Research Center for Functional Materials,
National Institute for Materials Science, 1-1 Namiki, Tsukuba 305-0044, Japan}
\author{T. Taniguchi} 
\affiliation{International Center for Materials Nanoarchitectonics,
National Institute for Materials Science, 1-1 Namiki, Tsukuba 305-0044, Japan}
\author{A. Vishwanath}
\affiliation{Department of Physics, Harvard University, Cambridge, MA~02138, USA}
\author{M. P. Zaletel}
\affiliation{Department of Physics, University of California, Berkeley, CA 94720, USA}
\affiliation{Materials Sciences Division, Lawrence Berkeley National Laboratory, Berkeley, CA 94720, USA}
\author{A. F. Young}   
\email{andrea@physics.ucsb.edu}
\affiliation{Department of Physics, University of California, Santa Barbara, CA 93106}
\begin{abstract}

\end{abstract}

\maketitle

\textbf{
At partial filling of a flat band, strong electronic interactions may favor gapped states harboring emergent topology with quantized Hall conductivity. 
Emergent topological states have been found in partially filled Landau levels~\cite{tsui_two-dimensional_1982} and Hofstadter bands~\cite{wang_evidence_2015, spanton_observation_2018}; in both cases, a large magnetic field is required to engineer the underlying flat band. The recent observation of quantum anomalous Hall effects~(QAH) in narrow band moir\'e systems~\cite{chen_tunable_2020,serlin_intrinsic_2020, stepanov_competing_2020, polshyn_electrical_2020} has led to the theoretical prediction that such phases may be realized even at zero magnetic field~\cite{zhang_nearly_2019, ledwith_fractional_2020, repellin_chern_2020, abouelkomsan_particle-hole_2020,wilhelm_interplay_2021}.
Here we report the experimental observation of insulators with Chern number $C=1$ in the zero magnetic field limit at $\nu=3/2$ and $7/2$ filling of the moir\'e superlattice unit cell in twisted monolayer-bilayer graphene (tMBG)~\cite{polshyn_electrical_2020,chen_electrically_2020, xu_tunable_2021, he_competing_2021}.
Our observation of Chern insulators at half-integer values of $\nu$ suggests spontaneous doubling of the superlattice unit cell, in addition to spin- and valley-ferromagnetism~\cite{kumar_generalizing_2014, wang_evidence_2015,spanton_observation_2018}. 
This is confirmed by Hartree-Fock calculations, which find a topological charge density wave ground state at half filling of the underlying $C=2$ band, in which the Berry curvature is evenly partitioned between occupied and unoccupied states. We find the translation symmetry breaking order parameter is evenly distributed across the entire folded superlattice Brillouin zone, suggesting that the system is in the flat band, strongly correlated limit. Our findings show that the interplay of quantum geometry and Coulomb interactions in moir\'e bands allows for topological phases at fractional superlattice filling that spontaneously break time-reversal symmetry, a prerequisite in pursuit of zero magnetic field phases harboring fractional statistics as elementary excitations or bound to lattice dislocations.}

\begin{figure*}[ht!]
\includegraphics[]{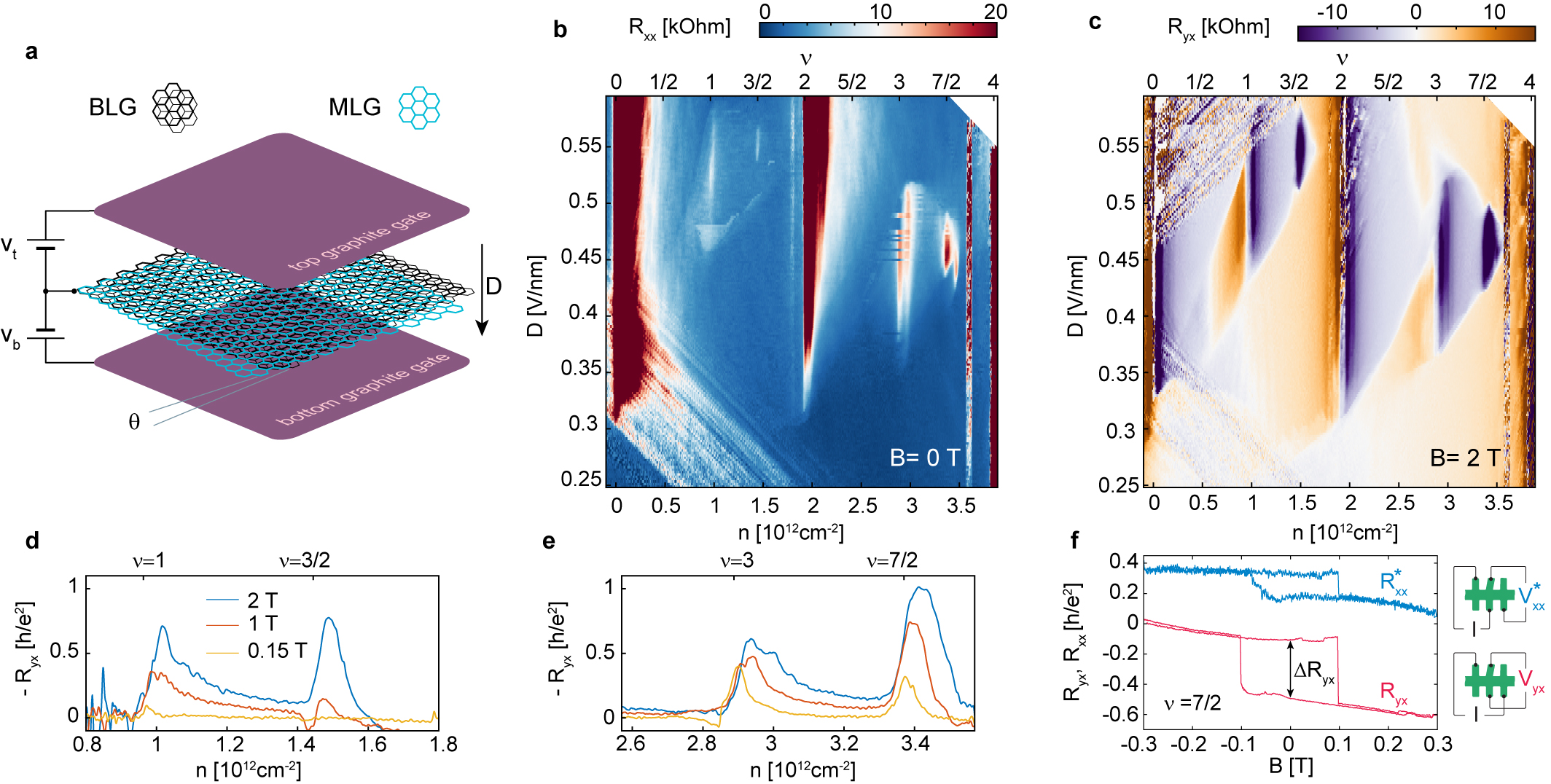} 
\caption{\textbf{Correlated states at half-integer filling of the moir\'e unit cell.}
\textbf{a,} Schematic of a tMBG device in which rotationally misaligned monolayer graphene (top) and Bernal-stacked bilayer graphene (bottom) are sandwiched between two graphite gates (hexagonal boron nitride (hBN) insulating spacer layers are not shown), enabling independent control of both the electric displacement field $D$ and the total carrier density $n$.  
\textbf{b, c}  Longitudinal resistance $R_{xx}$ and Hall resistance  $R_{yx}$ measured for a range of fillings spanning the conduction band at T=20~mK in device D2, with $\theta =1.29^\circ$. $R_{yx}$ is field-symmetrized at $B$=$\pm2$~T, while $R_{xx}$ is measured at zero magnetic field. In addition to well-developed states at $\nu$ =1,2 and 3, there are additional states at fillings $\nu=3/2$ and $7/2$, marked by sharp peaks in $R_{xx}$ and $R_{xy}$. 
\textbf{d, e} Hall resistance $R_{yx}$ measured at magnetic fields of 2, 1, and 0.15~T near $\nu$ =3/2 (d) and 7/2 (e) respectively. The displacement fields are  $D$=0.534 and 0.468~V/nm respectively. 
\textbf{f,}~Magnetic field dependence of $R_{yx}$ and $R^*_{xx}$  measured at $\nu$=7/2 and $D$= 0.456~V/nm, showing anomalous Hall effect with a hysteresis loop height of $\Delta R_{yx}\approx 0.4$~$h/e^2$. Right inset shows the measurement geometry. 
 }
\label{fig:1}
\end{figure*}

Broken time reversal symmetry is a requirement for many topological states of matter, including both those captured within an effectively noninteracting description~\cite{kitaev_periodic_2009} as well as those, like the fractional quantum Hall states, which are intrinsically correlation driven. 
While broken time reversal symmetry is traditionally achieved by the application of a large magnetic field, realizing topological phases at zero magnetic field has been a longstanding goal owing both to the increased experimental accessibility and to the possibility of enhanced energy scales.
A zero-field analog of the integer quantum Hall state was first theoretically proposed by Haldane~\cite{haldane_model_1988}, and later realized in magnetically doped topological insulators~\cite{chang_experimental_2013,deng_quantum_2020} and more recently in graphene based moir\'e heterostructures~\cite{chen_tunable_2020, serlin_intrinsic_2020}. In these systems, ferromagnetic interactions cause the two dimensional electrons to polarize into an odd-integer number of topological bands.  However, no topologically nontrivial zero-field phases have been experimentally demonstrated at fractional band filling, which would necessesitate a more complete rearrangement of single particle states by Coulomb interactions.  

Incompressible insulators in a two dimensional lattice can be classified by two quantum numbers, $C$ and $s$. $C$ is the net Chern number of the occupied states, and corresponds to the quantized  Hall conductivity $\sigma_{xy}=C e^2/h$, where $h$ is Planck's constant and $e$ is the electron charge. $s$ corresponds to the number of electrons bound to each lattice unit cell. In the limit of weak electronic interactions, $C$ and $s$ are both integers; however, when electron-electron interactions are sufficiently strong they may drive the onset of distinct gapped states with fractional $C$ and $s$. At high magnetic fields, graphene heterostructures realize all variants within this hierarchy.  Absent a moir\'e, high quality graphene heterostructures show integer and fractional quantum Hall states with $s=0$~\cite{dean_fractional_2020}. The addition of a moir\'e potential generates the fractal energy spectrum of the Hofstadter butterfly, which features gaps with integer $s$ and $C$ when an integer number of bands are filled~\cite{thouless_quantized_1982}.  When the bandwidth of the Hofstadter bands is tuned to be sufficiently small, additional gaps appear at fractional band filling. 
Symmetry broken Chern insulators (SBCIs) feature integer $C$ but fractional $s$. Their ground state breaks the superlattice symmetry, expanding the unit cell in a commensurate charge density wave pattern. Fractional Chern insulators (FCI) feature fractional $C$ and $s$, implying a fractionalization of charge of anyonic quasiparticles. 
Both SBCI and FCI states represent ground states incorporating correlations that reorder the filling of the single-particle states within a single Chern band. 

While topological states with integer quantized $s$ and $C$ have been observed at B=0, to date fractional $s$ or $C$ states have been observed only in the high $B$ limit. 
This failure is related to the difficulty of engineering bands that are simultaneously topologically nontrivial and have small bandwidth, so that the ground state will spontaneously break time reversal symmetry and in addition form either a charge density wave or topologically ordered state. 
Narrow bands engineered by introducing a moir\'e
superlattice in  graphene heterostructures provide an ideal venue to search for topological states with fractional quantum numbers. At the single particle level, these structures host valley-projected minibands with finite Chern number and small bandwidth.  Crucially, experiments have demonstrated that Coulomb interactions naturally favor spontaneous breaking of time reversal symmetry, manifesting as ferromagnetism~\cite{sharpe_emergent_2019, chen_tunable_2020-1} 
and the observation of quantum anomalous Hall effects~\cite{serlin_intrinsic_2020,polshyn_electrical_2020,chen_tunable_2020} at integer band filling $\nu=n/n_0$, where $n$ is areal carrier density, and $1/n_0$ is the area of a moir\'e unit cell (see Methods). These states--with integer $C$ and $s$---arise from  interaction-driven polarization of the electron system into an odd number of 
topologically nontrivial valley projected bands.  Moreover, evidence of correlation physics beyond this variety of spin- and valley ferromagnetism has recently emerged in the observation of incompressible states at fractional filling of moir\'e heterostructures composed of transition metal dichalcogenides, however the observed fractional $s$ states have $C=0$ ~\cite{regan_mott_2020,xu_correlated_2020,jin_stripe_2020, huang_correlated_2020}.

\section*{Results}
Here, we report the observation of symmetry broken Chern insulator states at half-filling of a Chern 2 band in twisted monolayer-bilayer graphene (tMBG) at zero magnetic field.
We study devices in which top and bottom graphite gates enable us to independently control both the total carrier density $n$ and the displacement field $D$ in the tMBG layer while simultaneously ensuring low charge disorder (see Fig.\ref{fig:1}a and Methods).
Previous studies of this system have demonstrated that the bandwidth and isolation of the low-energy bands are tuned by $D$~\cite{polshyn_electrical_2020, chen_electrically_2020, xu_tunable_2021, he_competing_2021}, with isolated flat bands realized for twist angles  $\theta\approx 0.8-1.4^\circ$ ~\cite{rademaker_topological_2020, ma_topological_2019, park_gate_2020}. 
We focus here on devices with $\theta \approx 1.3^\circ$ that show correlated insulating states at partial filling of the conduction band for a range of positive displacement fields near $D\approx0.45$~V/nm. Figure~\ref{fig:1}b shows the longitudinal resistance $R_{xx}$ in this regime measured at zero magnetic field and $T= 20$~mK  in device D2 with twist angle $\theta$ = $1.29^\circ$. 
Correlated states emerge at all integer fillings $\nu=1,2,3$ of the conduction band as shown in Fig.~\ref{fig:1}b,c. 
The state at $\nu=2$ has zero Chern number, while the states at $\nu=1$ and 3 are valley-polarized and characterized by Chern number $C$=2, as evidenced by measurements of quantum anomalous Hall effects ~\cite{polshyn_electrical_2020}. This finding is consistent with theoretical calculations of the underlying band structure in which valley-projected bands have $C=2$~\cite{rademaker_topological_2020, ma_topological_2019, park_gate_2020}. 
Adjacent to the $\nu$=1 and 3 insulators are regions marked  by high contrast in the Hall resistance (Fig.~\ref{fig:1}c and \ref{fig:S:FlavorSB}), which we take to correspond to fully flavor (spin and valley) symmetry broken metals.

Our primary finding is that correlated states also emerge at several half-integer fillings, manifesting as peaks in the longitudinal resistivity in Figs. \ref{fig:1}b at both $\nu=7/2$ and $\nu=3/2$.
At half-integer fillings we also observe pronounced features in the Hall resistance $R_{yx}$ (see Figs.~\ref{fig:1}c-e and Fig.~\ref{fig:S:Rxx_and_Rxy}).  
The $R_{yx}$ peak at $\nu$=7/2 approaches $h/e^2$ at 2~T and persists to low magnetic fields, showing hysteretic behavior with a jump of the Hall resistance of  $\approx 0.4 h/e^2$ at the coercive field (Fig.~\ref{fig:1}f). 
In contrast, at $\nu=3/2$ the peak in $R_{yx}$ disappears smoothly as the magnetic field is decreased, though a peak in $R_{xx}$ is still evident at $B=0$. 

\begin{figure}[ht!]
\includegraphics[]{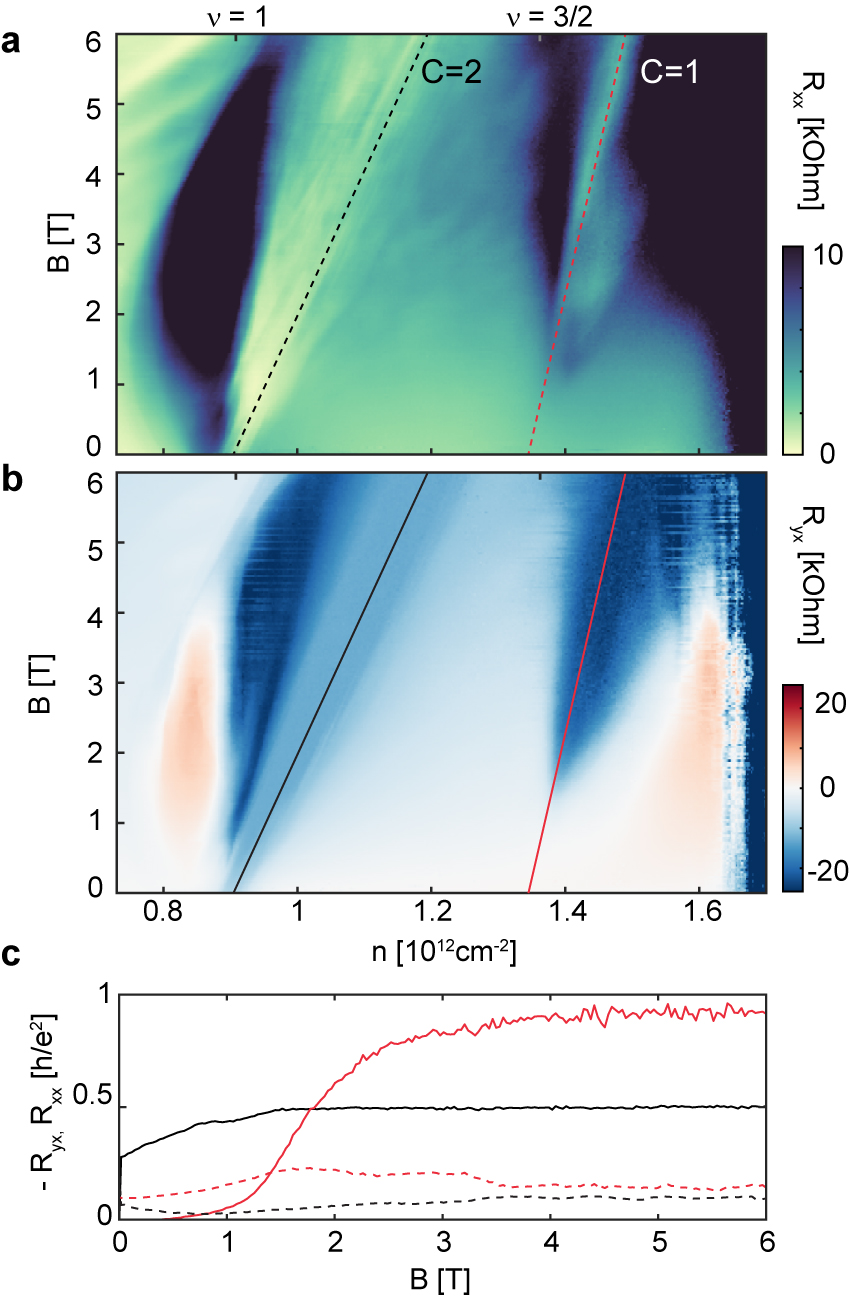} 
\caption{\textbf{Symmetry broken Chern insulator state at $\nu = 3/2$ in device D1, $\theta =1.25^\circ$.}
\textbf{a,b}, $R_{xx}$ and $R_{yx}$ measured at $D$=0.493~V/nm and $T$=20~mK show a quantum anomalous Hall state at $\nu=1$ and a Chern insulating state at 3/2 that appears at $B\approx 1$~T and becomes fully developed  above 2~T with $R_{xy}$ approaching $h/e^2$ and low $R_{xx}$. Lines originating from $\nu$= 1 and 3/2 have slopes expected~\cite{streda_quantised_1982} for states with Chern numbers of  $C=2$ and 1 respectively. 
\textbf{c}, The values of $R_{xx}$ and $R_{yx}$ along the lines shown in (a,b). 
}
\label{fig:2}
\end{figure}

The large Hall effect and commensurate fractional filling suggest that new Chern insulator states emerge at $\nu=3/2$ and 7/2, with the former requiring a small magnetic field for stability. To further investigate this hypothesis, we study the  magnetic field dependence of these states. 
Fig.~\ref{fig:2} shows $R_{xx}$ and $R_{yx}$ as a function of magnetic field for a range of filling that includes both $\nu$=1 and 3/2, acquired for a second device (D1, with  $\theta =1.25^\circ$). 
For both states, the density corresponding to the Hall resistivity maximum and longitudinal resistivity minimum evolve with magnetic field following a linear trajectory, as expected from the Streda formula, $C=(h/e)\partial n/\partial B$, which relates the charge carrier density of an incompressible state with applied magnetic field~\cite{streda_quantised_1982}.
The Streda formula captures the fact that the degeneracy of the occupied Chern bands is sensitive to the magnetic flux, with a coefficient that encodes the Chern number $C$ of the topological gap~\cite{xiao_berry_2010}. Comparing data from the $\nu=1$ and $3/2$ Chern insulators suggests Chern numbers of $2$ and $1$, respectively.  This is confirmed by the magnetic field evolution of the Hall resistance, which approaches the corresponding quantized values for both states  for $B>2$~T~(Fig.~\ref{fig:2}c). In both states, the longitudinal resistance is also comparatively small, and shows a minimum as a function of magnetic field and carrier density as expected for a Chern insulator dominated by chiral edge-state conduction. 
In contrast to the $\nu$=1 state, which persists down to zero field and exhibits a quantum anomalous Hall effect, the state  at $\nu$=3/2 only emerges in finite magnetic field (see also Fig.~\ref{fig:S:D1ThreeHalves}). 
Similar behavior is observed in device D2 (Fig. \ref{fig:S:Rxx_and_Rxy}).

\begin{figure*}[ht!]
\includegraphics[]{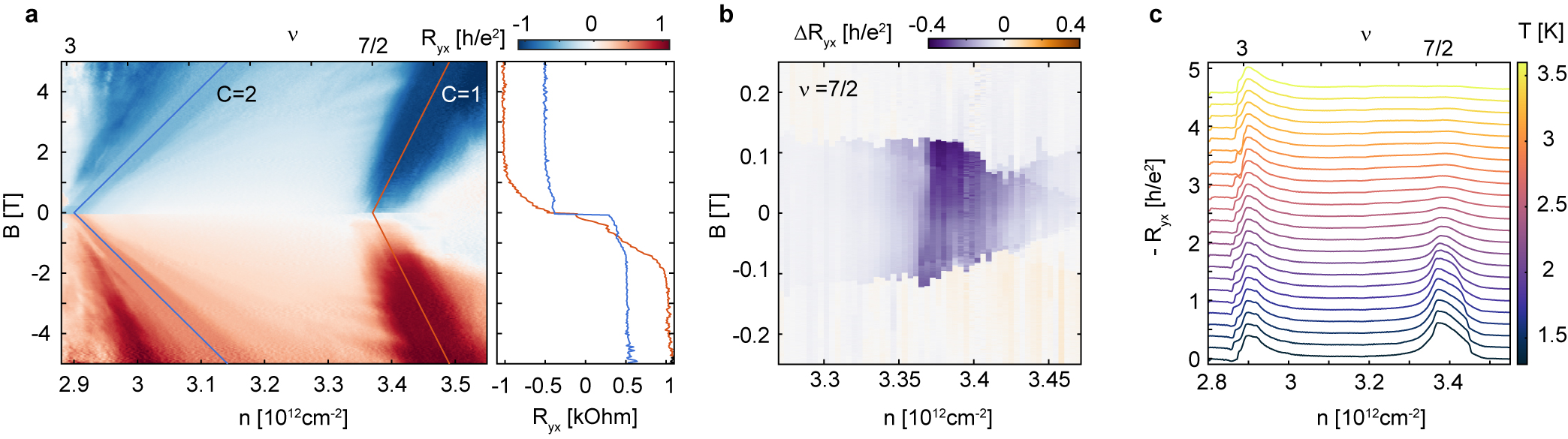} 
\caption{\textbf{Evidence for non-trivial topology and ferromagnetism at $\nu = 7/2$.}
\textbf{a,} Low-field evolution of Hall resistance, measured at $D=0.466$~V/nm for a range of $n$ spanning states at $\nu$ =3 and 7/2  in device D2. Lines originating from $\nu$ =3 and 7/2 are guides for the eye that have slopes expected for Chern insulating states with $C=2$ and 1 respectively. The inset shows $R_{yx}$ plotted along the lines in the main panel, which saturate to approximately $0.5h/e^2$ and $h/e^2$. At zero magnetic field both traces exhibit finite jumps.
\textbf{b,} Doping dependence of the magnetic hysteresis $\Delta R_{yx}$, as defined in Fig.~\ref{fig:1}f, in the vicinity of $\nu$=7/2 measured at $D=0.456$~V/nm.
\textbf{c,} Temperature dependence of the Hall resistance measured at $B=0.25$~T and $D=0.451$~V/nm. The $R_{yx}$ peak at $\nu$=7/2 disappears near 2.5~K.}
\label{fig:3}
\end{figure*}

The magnetic field evolution of a two dimensional gapped state characterized by quantum numbers ($C$, $s$) is described by the Diophantine equation $\nu = C n_\Phi+s$, where $n_{\Phi}$ is the number of magnetic flux quanta threading each moir\'e unit cell~\cite{wannier_result_1978,streda_quantised_1982}.
From the observed magnetic field dependence of the $\nu=3/2$ state we determine $(C,s) = (1,3/2)$. 
A half-integer value of $s$ can be realized only if electronic interactions allow for a fraction of a charge to be bound to a unit cell~\cite{oshikawa_commensurability_2000}.
This can occur in one of two ways.
First, the system may support excitations with fractional charge $e/2$, as in the fractional quantum Hall effect; however, in the likely realizations of this scenario the Chern number would also be fractional. 
More conservatively, electron interactions may lead the ground state to spontaneously break the original moir\'e superlattice symmetries, so that an integer number of electrons are bound to a doubled unit cell; the absence of charge fractionalization however mandates an integer quantized Hall conductivity \cite{kumar_generalizing_2014}.
We attribute the $\nu=3/2$ state to just such a symmetry broken Chern insulator. 
This SBCI state splits the spin- and  valley- polarized  Chern 2 band into two Chern 1 bands, and exists only in the region of the $(n,D)$ plane near the $\nu=1$ state, where both spin- and valley- degeneracies are fully broken (see Fig.~1c,~\ref{fig:S:D1ThreeHalves}, ~\ref{fig:S:FlavorSB}).  

The state observed at $\nu=7/2$ shares many features with that at 3/2. It is a symmetry-broken Chern insulating state with $C$ =1 and $s$=7/2 as follows from both the Hall conductivity and the $n$-$B$ dependence  (Fig.~\ref{fig:3}a). However, in contrast to $\nu=3/2$, the 7/2 state persists all the way down to zero magnetic field  and exhibits a large anomalous Hall effect with magnetic hysteresis of $\Delta R_{yx}\approx 0.4 h/e^2$ and coercive fields of $\approx$100~mT (Fig.~\ref{fig:3}b and Fig.~\ref{fig:S:Hysteresis}). While the Hall conductivity is not perfectly quantized, multiterminal transport measurements strongly suggest that transport is dominated by chiral edge states at $B=0$, with imperfection of the quantization arising from structural inhomogeneity within the device~\cite{uri_mapping_2019,tschirhart_imaging_2020} (see Fig.~\ref{fig:S:D2ChernInsulator}). 
Taking the temperature at which the  anomalous Hall  resistance at $\nu$=7/2   disappears as an estimate of the Curie temperature we obtain  $\approx2.5$~K (Fig.~\ref{fig:3}c). This is only about a factor of two smaller than the Curie temperature of QAH states at $\nu$=1 and 3~\cite{polshyn_electrical_2020}. 

\begin{figure}
    \includegraphics{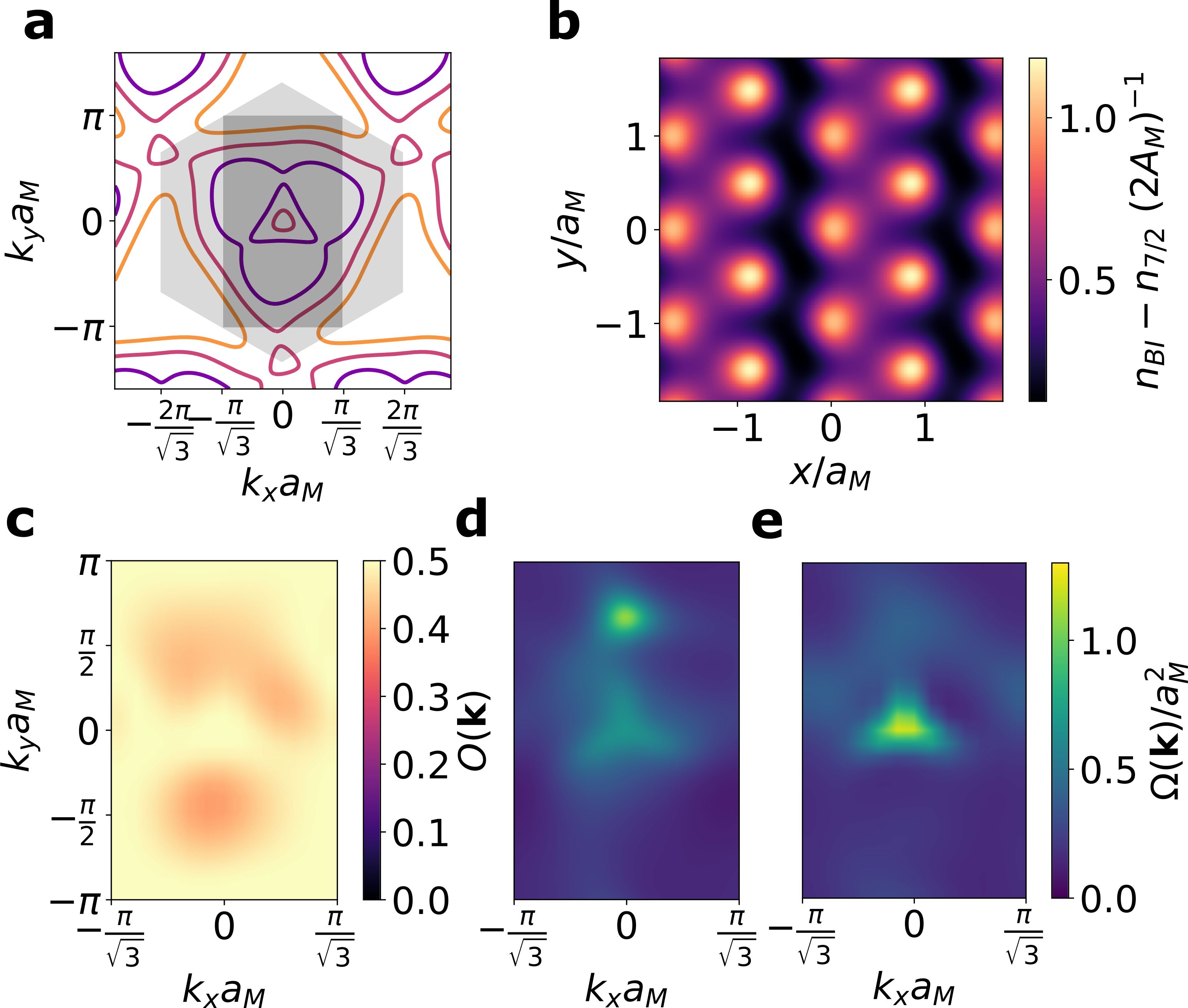}
    \caption{\textbf{Hartree-Fock calculation of $\nu = 7/2$ SBCI}. \textbf{a}, Contour plot of the non-interacting Fermi surfaces of the $C=2$ band in graphene valley $K$. The purple, red, and yellow lines correspond to quarter, half, and three quarters filling of the band respectively. The lightly shaded hexgagon and the darker shaded rectangle are the Brillouin zones before and after translation symmetry breaking. \textbf{b}, Density of the unfilled $C=1$ band after translation breaking, measured by the difference between the density of the $\nu = 4$ band insulator and the $\nu = 7/2$ SBCI. \textbf{c}, Translation breaking order parameter is close to $1/2$ everywhere in the Brillouin zone similar to a strong coupling sublattice polarized state. \textbf{d}, Half of the non-interacting Berry curvature, folded onto the translation breaking Brillouin zone. \textbf{e}, Berry curvature for the $C=1$ unfilled band in the $\nu = 7/2$ Hartree-Fock SBCI ground state.
    }
    \label{fig:theory}
\end{figure}

Calculations of the tMBG band structure illuminate the essential features of the moir\'e minibands that enable SBCI states at fractional filling. 
In addition to the bandwidth $W<10$~meV, which is significantly smaller than the scale of the Coulomb interaction $E_C\approx30$~meV, band structure calculations show a relatively homogeneous distribution of the Berry curvature for the conduction band~\cite{park_gate_2020, rademaker_topological_2020}, with similar contributions arising from the neighborhoods of both the $\Gamma$ and $K$ points within the moir\'e Brillouin zone. 
Interaction driven doubling of the unit cell at half filling of the $C=2$ band is thus likely to equally partition the evenly distributed Berry curvature, yielding a $C=1$ gap. 
Such symmetry breaking states constitute a lattice generalisation of quantum Hall ferromagnetism~\cite{kumar_generalizing_2014}.

A Hartree-Fock calculation (see Fig.~\ref{fig:theory}), generalized to include translation symmetry breaking that doubles the unit cell, obtains a SBCI ground state for a range of displacement fields around $0.4$ V/nm at filling $\nu = 7/2$.  The unfilled translation-breaking band is spin and valley polarized and has $C = 1$; its density and Berry curvature are shown in Fig.~\ref{fig:theory}b,e. The Berry curvature does not have a visible peak at the $K$ point, unlike the non-interacting Berry curvature, likely due to interaction-induced mixing with the $C=-1$ band below. 
Finally, the calculated density confirms the stripe order that is inferred from the experimentally-observed half-integer filling, which, in the absence of fractionally quantized Hall resistance, strongly implies unit cell doubling. 

The Hartree-Fock state we obtain is very similar to the strong coupling idealized model of sublattice polarization discussed in~\cite{kumar_generalizing_2014}. In particular, the translation breaking order parameter for a fully sublattice polarized state on a square lattice is $O(\mathbf{k}) = \langle c^\dag_{\mathbf{k}} c_{\mathbf{k} + \mathbf{Q}} \rangle = 1/2$. The analogous order parameter for monolayer-bilayer graphene is plotted in Fig.~\ref{fig:theory}c; it is close to $1/2$ everywhere in the Brillouin zone. This is in sharp contrast to weak coupling translation symmetry breaking, where the order parameter is only significant in regions where the Fermi surface folds on top of itself, such as at the zone boundaries.

\section*{Discussion}
In their underlying symmetry and topology, the SBCI states we report here belong to the same universality class as those reported previously at high magnetic fields in partially filled Hofstadter bands~\cite{wang_evidence_2015, spanton_observation_2018, saito_hofstadter_2021}, where they typically appear at magnetic fields of at least 8 Tesla.  Charge density wave physics may also underlie states observed in twisted bilayer graphene at low magnetic fields where $C+s$ is an odd integer\cite{yankowitz_tuning_2019,stepanov_competing_2020,pierce_unconventional_2021}.  Charge density wave order can potentially explain these features~\cite{kang_non-abelian_2020,pierce_unconventional_2021} although other mechanisms such as mixing with remote bands~\cite{xie_nature_2020} have also been invoked. In comparison, the insulating features described here are more directly interpreted as charge density waves, given that they appear at half-integer filling. 

In all cases, interactions gap the Fermi surface by breaking the superlattice symmetry at fractional filling of a Chern band with $C\geq2$. 
SBCI states arise when the Berry curvature is evenly partitioned between occupied and unoccupied states at fractional filling, leading to a topological gap with integer Chern number intermediate between those of adjacent gaps at integer band filling.  However, the SBCI states that we report here differ in the key respect that time reversal symmetry is broken spontaneously rather than through the application of a magnetic field.  In all previous realizations of correlated topological states at fractional band filling--which includes both SBCIs but also fractionalized states such as fractional quantum Hall effects~\cite{tsui_two-dimensional_1982} and fractional Chern insulators in Hofstadter bands~\cite{spanton_observation_2018}---the underlying Chern band structure is itself a product of the applied magnetic flux.  As a result, these systems do not provide a path way to zero magnetic field realizations. In contrast, in twisted monolayer-bilayer graphene the Chern band structure is intrinsic to the moir\'e miniband structure without applied flux.  

In contrast to conventional charge density waves that are well described by a classical order parameter, charge density waves that split a C=2 band have certain special topological features.  In these states, topology and charge-density-wave order become entwined.  A Chern two band can be understood as analogous to a two-component quantum Hall system in which sub-lattice plays the role of pseudo-spin~\cite{kumar_generalizing_2014}, so that dislocations in the charge density-wave order map to  non-Abelian fluxes in the pseudo-spin space~\cite{barkeshli_topological_2012}. In the SBCI states we find here, this leads to an entwining of charge with topological defects such as lattice dislocations. Directly probing these novel features, and associated phenomena and functionalities, is an exciting and accessible goal for the future. 
Our results suggest that such states and their attendant defects should be readily accessible in the vast design space of flat band moire heterostructures. 

\section*{Methods}
\subsection*{Device fabrication}
Twisted monolayer-bilayer graphene devices  were fabricated using a dry transfer method. A PDMS stamp covered with a thin PC (polycarbonate) film was used to subsequently pick up flakes and to create the following van der Waals hetersostructures (from top to bottom): hBN - FLG - hBN - MLG - BLG - hBN -FLG, where hBN is hexagonal boron nitride (usually 30-50~nm thick), FLG - few layer graphene, MLG - monolayer graphene, and BLG - Bernal-stacked bilayer graphene. 
The vdW heterostructures were further processed using e-beam lithography, $\mathrm{CHF_3/O_2}$ etching, and edge contacts deposition (Cr/Pd/Au with thicknesses of 1.5nm/15nm/250nm). 
Devices D1 and D2 (see fig.~\ref{fig:S:Devices}) in this study are the same as in Ref.~\cite{polshyn_electrical_2020}, which contains more details about the fabrication process. 

\subsection*{Device characterization}
 Top and bottom graphite gates allow us to tune both the total carrier density, $n=c_t\mathrm{v}_t+c_b \mathrm{v}_b$, and the electric displacement field, $D=\left(c_t\mathrm{v}_t-c_b\mathrm{v}_b\right)/2\varepsilon_0$. Here  $\mathrm{v}_{t(b)}$ is the applied voltage and $c_{t(b)}$ is the capacitance per unit area of the top (bottom) gate, $\varepsilon_0$ is the  vacuum permittivity. $c_t$ and  $c_b$ are determined by fitting the features of Landau fans for the devices.

A rotational misalignment between MLG and BLG layers by angle $\theta$ produces a moir\'e superlattice with  unit cell area of $1/n_0\approx\sqrt{3}a^2/(2 \sin^2\theta)$, where $a=2.46$~\AA. We determine the twist angles of devices by matching the position of the correlated insulating states at integer fillings. In device D2, the positions of features in $R_{yx}$ is fitted well with twist angle $\theta$ = $1.29^\circ$  which corresponds to $n_0$=0.963~$\mathrm{\times10^{12}cm^{-2}}$.
In device D1, we obtain $\theta=1.25^\circ$ and $n_0$=0.908~$\mathrm{\times10^{12}cm^{-2}}$.

Unless specified otherwise, the measurements were performed in a cryogen-free dilution refrigerator at nominal temperature of approximately $20$~mK. In case of Fig.~\ref{fig:S:FlavorSB}c, the measurements were done in a wet variable-temperature insert with the sample in helium vapour.  Transport measurements were done using lock-in amplifiers (SRS), voltage pre-amplifiers (SRS), and  a current amplifier (DL Instruments) with   excitation currents of 1-5~nA at 17.77~Hz.

In device D2, unless indicated otherwise, the four point resistance measurements of $R_{yx}$ and $R_{xx}$ were done by measuring voltage drop across the pairs of contacts labeled as (b, e) and (f, e) respectively as indicated in   Fig.~\ref{fig:S:Devices}b.

\subsection*{Hartree-Fock simulations}
Hartree-Fock simulations were performed with the non-interacting Hamiltonian identical to the Hamiltonian described in the supplement of Ref.~\cite{polshyn_electrical_2020}. We used the parameters $\theta = 1.29^\circ$, $D = 0.4$ V/nm, $w_{AA}/w_{BB} = 0.75$ in all plots. We allowed for translation symmetry breaking that doubled the unit cell; translation breaking states that more than doubled the unit cell or broke out of plane spin rotations were excluded. More details and results are shown in the supplemental material.

\section*{acknowledgments}
The authors are grateful to Jihang Zhu for fruitful discussions. 
AFY acknowledges support of the Office of Naval Research under award N00014-20-1-2609, and the Gordon and Betty Moore Foundation under award GBMF9471. 
MPZ acknowledges support of the ARO under MURI W911NF-16-1-0361.  
KW and TT acknowledge support from the Elemental Strategy Initiative conducted by the MEXT, Japan, Grant Number JPMXP0112101001, JSPS KAKENHI Grant Numbers JP20H00354 and the CREST(JPMJCR15F3), JST.  AV was supported by a Simons Investigator Award. 
P.L. was supported by the Department of Defense (DoD) through the National Defense Science \& Engineering Graduate Fellowship (NDSEG) Program.

\section*{Author contributions}
HP, MAK, and YZ fabricated the devices.  HP performed the measurements, advised by AFY.  KW and TT grew the hexagonal boron nitride crystals.  TS, PL, MPZ, and AV contributed to the theoretical interpretation and performed the Hartree-Fock calculations.  HP, PL, and AFY wrote the manuscript with input from all other authors.

\section*{Competing interests}
The authors declare no competing interests.

\section*{Data availability}
The data that support the plots within this paper and other findings of this study are available from the corresponding author upon reasonable request.

%


\renewcommand{\figurename}{\textbf{Extended Data Fig.}}
\renewcommand{\thefigure}{E\arabic{figure}}
\setcounter{figure}{0}

\begin{figure*}[ht!]
\includegraphics[]{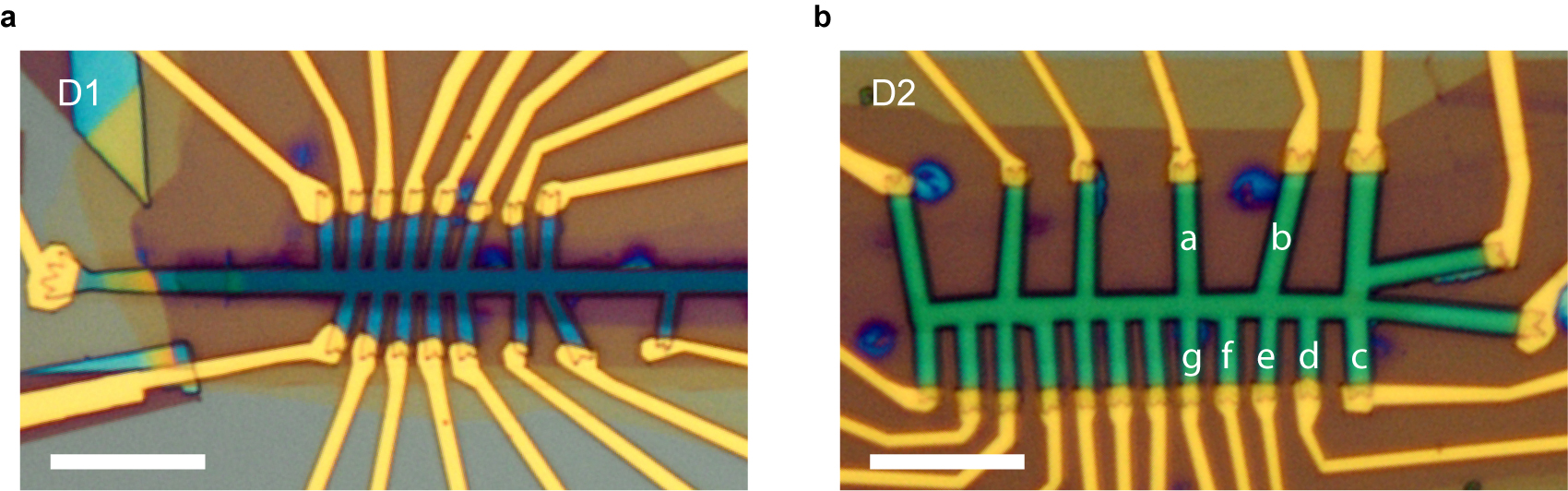} 
\caption{\textbf{tMBG devices.} Optical images of tMBG devices D1 (a) and D2 (b). Scale bars are 10~$\mathrm{\mu m}$.}
\label{fig:S:Devices}
\end{figure*}

\begin{figure*}[ht!]
\includegraphics[]{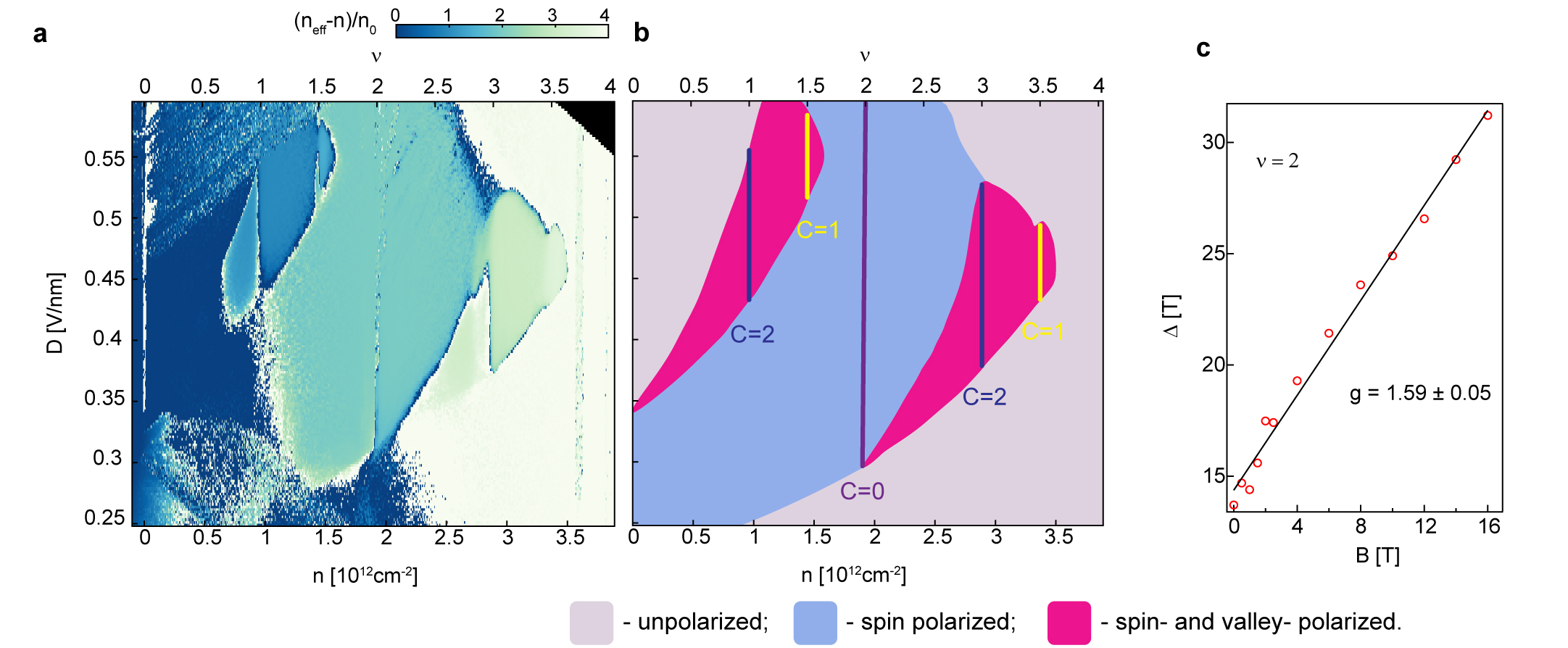} 
\caption{\textbf{Spin- and valley- symmetry breaking.}
\textbf{a}, Metallic phases with broken flavor (spin and valley) degeneracies can be highlighted using quantity $n^*=(n_{\mathrm{eff}}-n)/n_0$, where $n_{\mathrm{eff}}$ is the effective  carrier density which was extracted here from $R_{yx}$ measured at $B$ = 1~T in device D2. $n^*$ corresponds to the filling of the superlattice at which the band  that is being filled becomes completely empty (full) for    n-type (p-type) Fermi surface. Thus, integer values of $n^*$ that are intermediate between 0 and 4 indicate interaction-induced symmetry breaking of the flavor degeneracies. 
\textbf{b}, Diagram of flavor symmetry broken phases corresponding to (a). The boundaries were determined using $n^*$ as well as measured $R_{xx}$ and $R_{yx}$. The incompresssible  states at commensurate fillings are labeled with the Chern numbers observed for $B>0$.
The blue region around the $\nu$=2 insulating state corresponds to a two-fold degenerate state, which we  identify as spin-polarized phase because of the spin polarization of $\nu=2$ (see (c)) and the absence of the anomalous Hall effect~  \cite{polshyn_electrical_2020}. Smaller red regions around $\nu$=1 and 3 states  correspond to phases with fully lifted  spin-and valley- degeneracies,  in which a single $C=2$ band is being filled. States at $\nu=3/2$ and $7/2$ emerge within these regions. 
\textbf{c}, In-plane magnetic field dependence of the gap at $\nu$=2 determined from the thermal activation of $R_{xx}$ in device D1 at $D$=0.443~V/nm. Linear fit yields $g$-factor of $1.59\pm0.05$ which indicates large spin polarization. The determined value is lower than $g=2$ which  is expected for fully spin-polarized state. The exact origin of this discrepancy is not fully understood and could include  orbital effects of the in-plane magnetic field.
}
\label{fig:S:FlavorSB}
\end{figure*}

\begin{figure*}[ht!]
\includegraphics[]{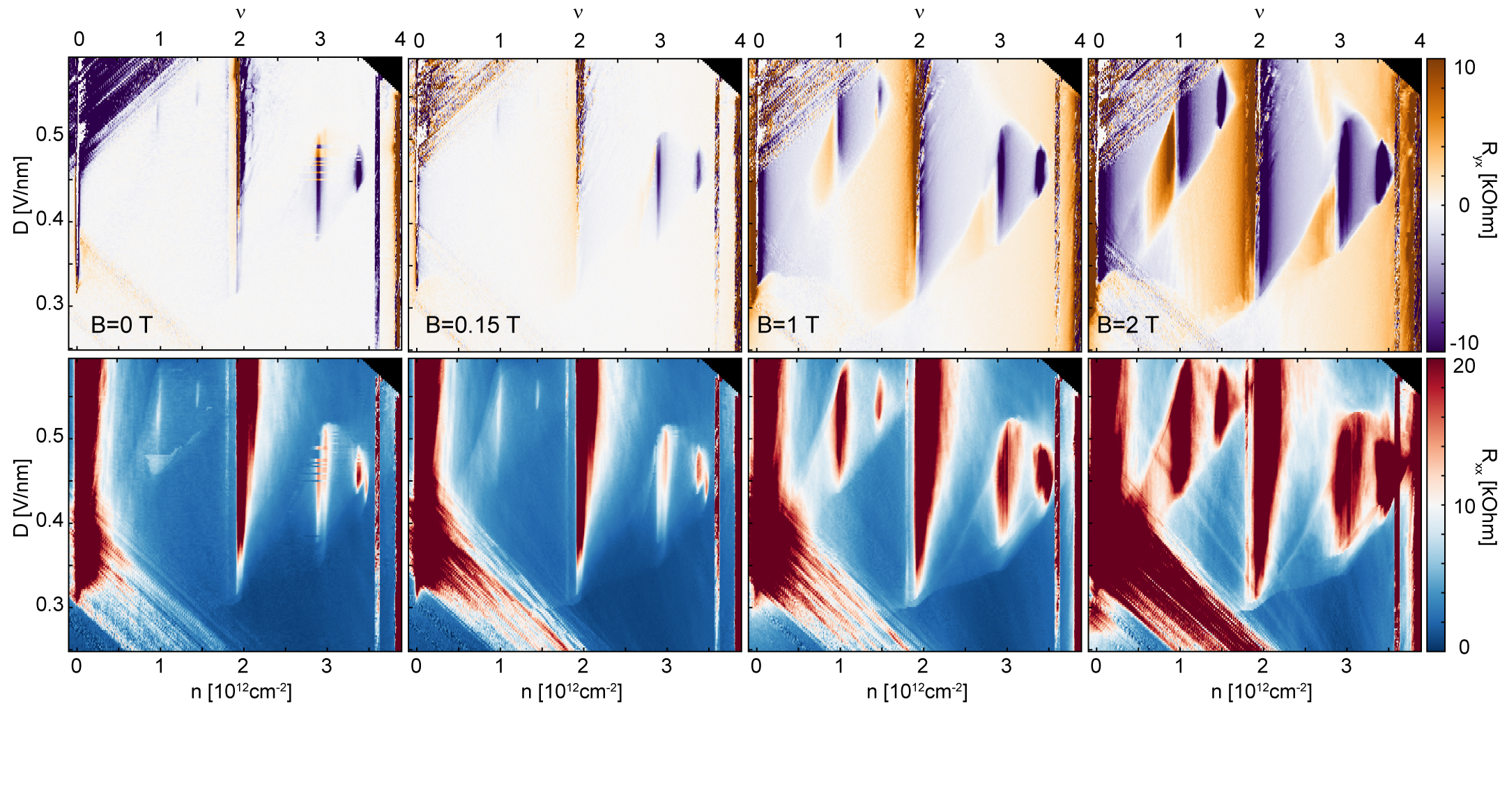} 
\caption{\textbf{$\mathbf{R_{yx}}$ and $\mathbf{R_{xx}}$ measured in device D2 at selected magnetic fields.} Magnetic fields are indicated in panels. The $R_{xx}$ ($R_{yx}$) data taken at finite fields is symmetrized (antisymmetrized) with respect to the magnetic field reversal. }
\label{fig:S:Rxx_and_Rxy}
\end{figure*}

\begin{figure*}[ht!]
\includegraphics[]{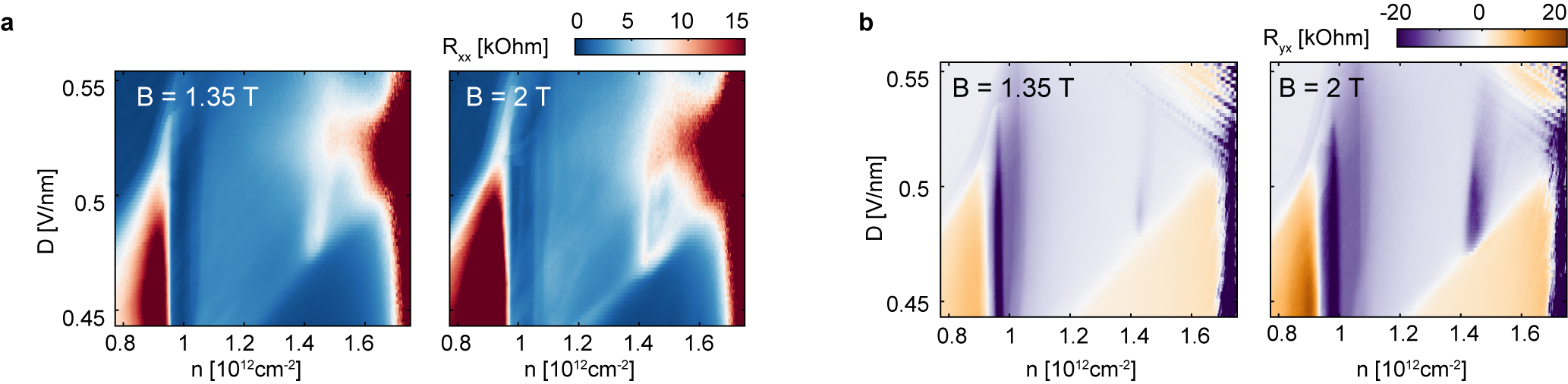} 
\caption{\textbf{Emergence of the SBCI state at $\nu$ =3/2 in device D1. }
\textbf{a,~b}, $R_{xx}$ (a) and $R_{yx}$ (b) measured in the region around  $\nu$ = 3/2 state at $B=1.35$ and 2~T. At  B=1.35~T, weak peaks in  $R_{xx}$ and $R_{yx}$ emerge at $\nu$=3/2. At 2 T, the peak in $R_{yx}$ is enhanced while $R_{xx}$ shows a dip developing at $\nu$=3/2.
}
\label{fig:S:D1ThreeHalves}
\end{figure*}

\begin{figure*}[ht!]
\includegraphics[]{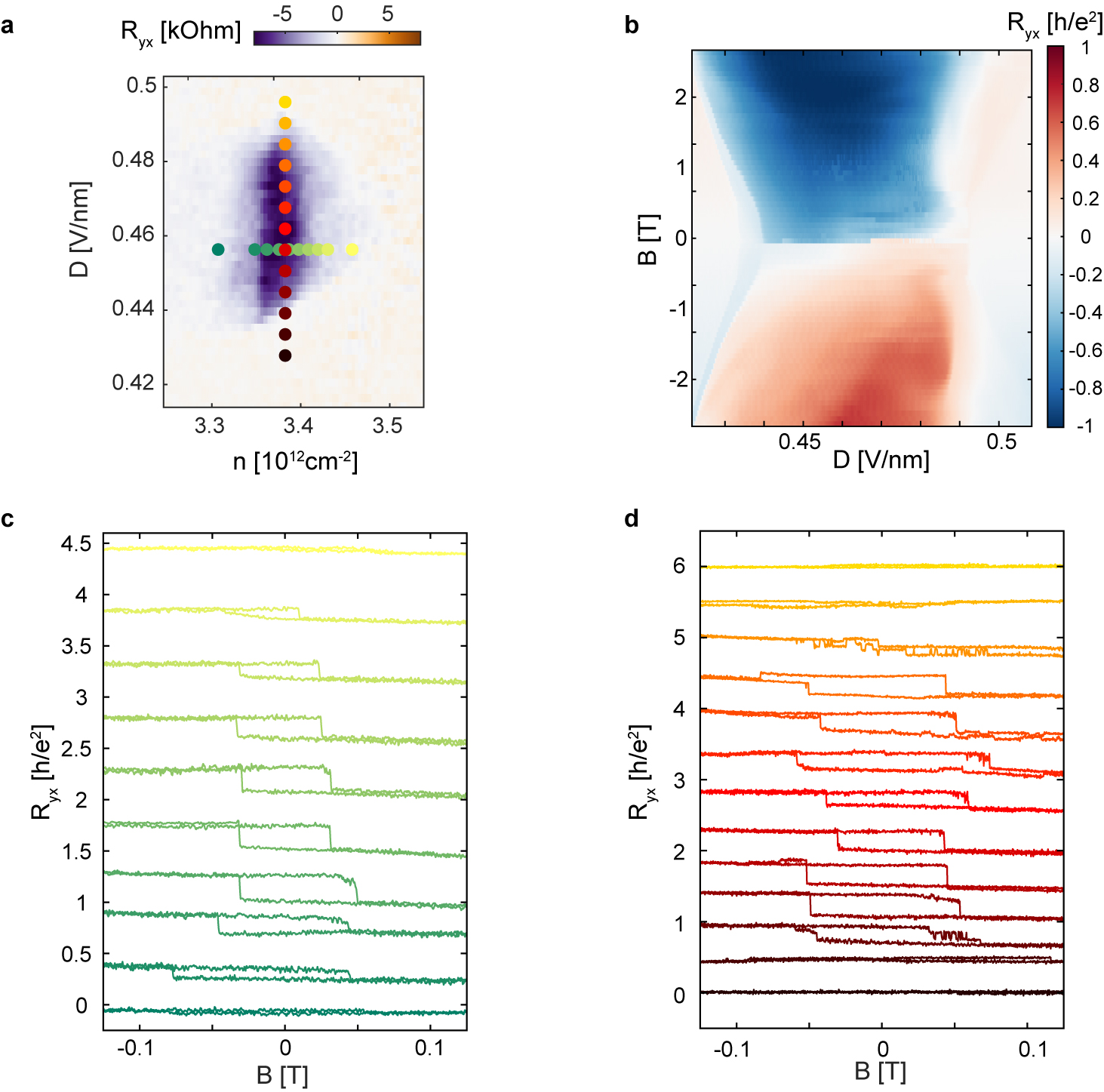} 
\caption{\textbf{Dependence of the anomalous Hall effect on $\mathbf{n}$ and $\mathbf{D}$ near $\nu$ = 7/2 in D2. 
}
\textbf{a}, $R_{yx}$ in the vicinity of a $\nu$=7/2 state measured at $B=$0.15~T.
\textbf{b}, $D$-field evolution of $R_{yx}$ measured at $n$= $3.39\times 10^{12}$~$\mathrm{cm^{-2}}$. In this measurement, the fast sweep axis is $D$. The range of $D$ at which the SBCI state is observed increases with $B$, indicating that  the magnetic field stabilizes the state.
\textbf{c},\textbf{d}, The dependence of the magnetic hysteresis on $n$~(c) and $D$ (d). The values of $(n,D)$  for each hysteresis loop are indicated by markers of the same color in (a). The curves are offset by $0.5 h/e^2$ for clarity.
}
\label{fig:S:Hysteresis}
\end{figure*}

\begin{figure*}[ht!]
\includegraphics[]{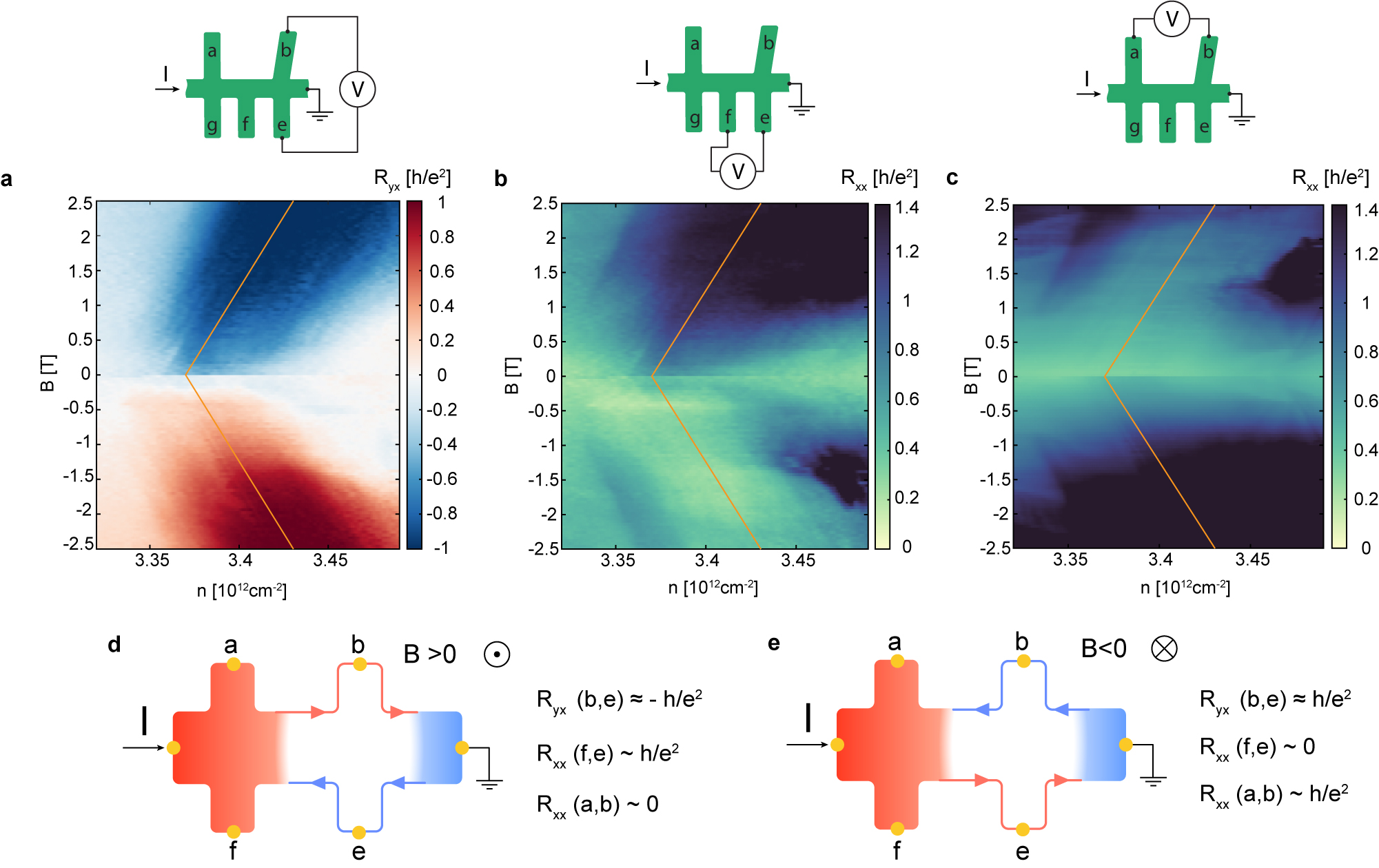} 
\caption{\textbf{Details of the magnetic field evolution of $R_{xx}$ (a) and $R_{yx}$ (b,c) near $\nu=7/2$ in D2.}
\textbf{a,b,c} $R_{yx}$ and $R_{xx}$  measured at $D$ = 0.466~V/nm. The corresponding contact configurations are shown in the diagrams above the plots.
 $R_{xx}$ in (b) and (c) were measured on the opposite sides of the device. The tilted lines are guides for the eye and have the slope expected for $C$=$\pm1$ states. A number of fine features in (a-c), some of which persist down to zero field, align perfectly with  these lines, providing further evidence for QAHE state at  $\nu$=7/2 which is deteriorated by the disorder. We speculate that the finite values of $R_{xx}$ originate from the twist angle disorder which results in the presence of metallic regions in the device and hence interferes with the edge state transport. It is worth pointing out an interesting feature of $R_{xx}$: $R_{xx}$ is low for $B<0$, but becomes of the order of $h/e^2$ for $B>0$, as shown in (b). This trend reverses on the opposite side of the device, as shown in (c). A plausible scenario that can produce this behaviour is illustrated in \textbf{d} and \textbf{e}. Let us consider a situation when  the  device region between contacts "b" and "e" is insulating in the bulk (shown in white) and has fully-developed edge states, while the adjacent regions remain metallic (shown in red and blue) due to the local twist angle variation. The color marks local electrical potential, increasing from blue to red. As a result of the edge state transport, there is a significant voltage drop $\approx I h/e^2$ across the insulating region. In contrast to this, the electrical potential changes only gradually within the metallic regions since they have resistivity which is much smaller than $h/e^2$. Because of the chiral nature of the edge states, $R_{xx}\sim h/e^2$ when measured on the side of the device where the "cold" edge states equilibrates with the metallic region, while $R_{xx}\sim 0$ on the other side of the device. We note that in this scenario, $R_{yx}$ measured using contacts "b" and "e" is quantized.}
\label{fig:S:D2ChernInsulator}
\end{figure*}

\newpage\clearpage
\renewcommand{\figurename}{\textbf{Supplementary Fig.}}
\renewcommand{\thefigure}{S\arabic{figure}}
\setcounter{figure}{0}

\widetext
\section{Supplementary Information}

The non-interacting band structure was obtained from the Hamiltonian identical to those described in \cite{polshyn_large_2019}. We obtain the renormalized non-interacting band structure using subtraction method described in the Appendix of \cite{soejima_efficient_2020}, and perform Hartree-Fock calculation with respect to double gate screened Coulomb potential $V_q = \tanh(dq)[2\epsilon_0 \epsilon_r q]^{-1} $ with relative permittivity $\epsilon_r = 20$ and gate distance $d = 120$ nm. 

Hartree-Fock simulations were performed for a range of parameters close to the filling $\nu = 7/2$. We allowed for translation symmetry breaking states that doubled the unit cell by demanding that translations along the $y$ axis are preserved, and that other translations are preserved at order two. This leads to the rectangular Brillouin zone shown in Figure \ref{fig:theory}a. While there are two other ways to double the unit cell, they are related to this one by $C_3$ symmetry. We do not allow for spin rotations about the out of plane axis to be broken; while this still allows for spin polarized states it excludes some spin density wave states. We confirm the Chern number of the Hartree-Fock ground state via the plaquette method \cite{fukui_chern_2005} and via winding of Wannier polarization, as described below.

A crude sweep across different displacement fields found an SBCI ground state close to $D = 0.4$ V/nm. This state has $C=-1$, as it leaves a single $C=1$ band unfilled, and it breaks translation symmetry with close to maximal magnitude across the entire Brillouin zone as shown by the plot of the translation breaking order parameter in Figure \ref{fig:theory}c. We chose
\begin{equation}
    O (\mathbf k) = \sqrt{ \sum_{n,n'} | \langle c^\dag_{\tau s n \mathbf k} c_{\tau s n' \mathbf k + \mathbf Q} \rangle |^2 },
\end{equation}
where $\tau$ and $s$ label the valley and spin that are not fully filled and $n$ and $n'$ index the non-interacting bands; most of the contribution is from the term where $n$ and $n'$ both index the $C=2$ band as expected. The translation breaking wavevector is $\mathbf{Q} = \frac{2\pi}{\sqrt{3}} \hat{\mathbf{x}}$.

\begin{figure}[ht!]
    \centering
    \includegraphics[width=\textwidth]{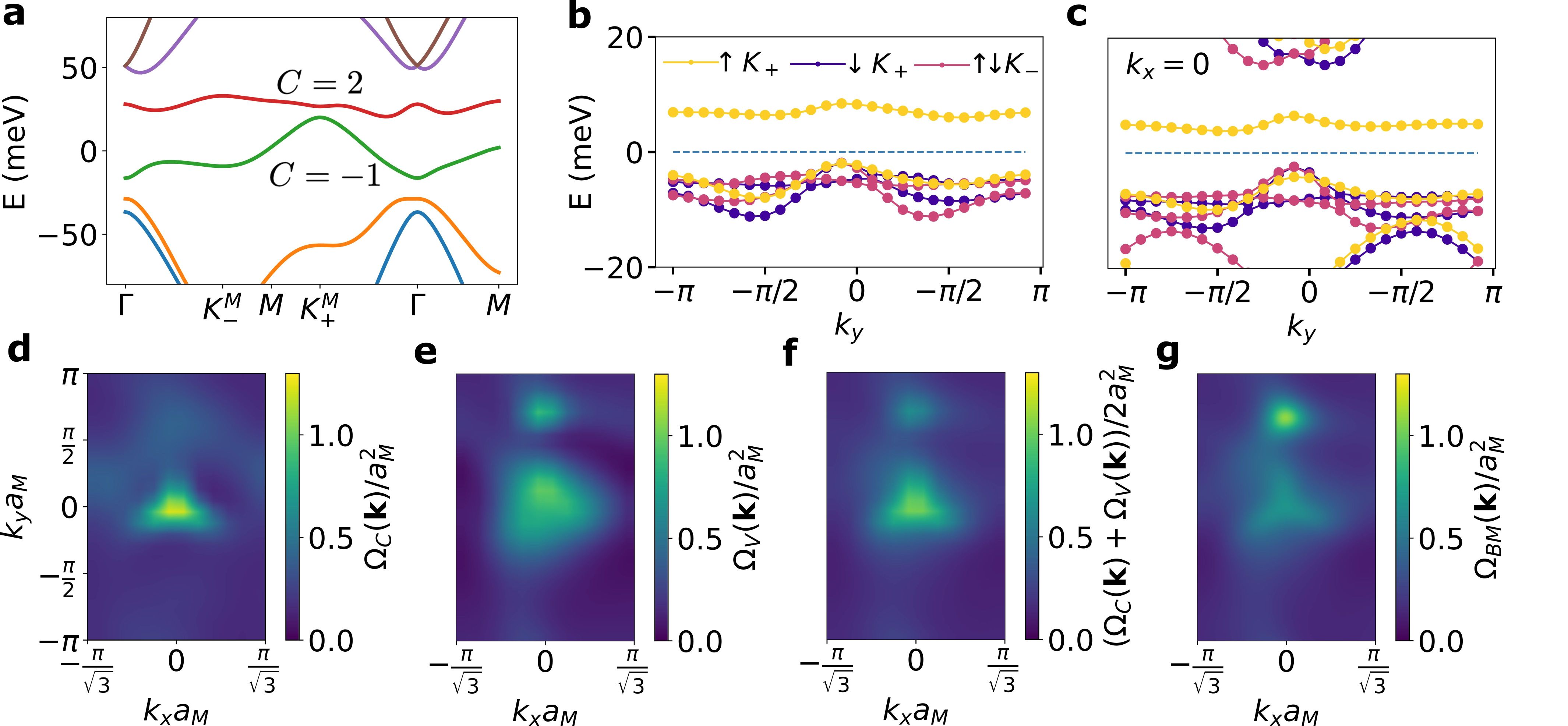}
    \caption{
    \textbf{Band structures and Berry curvatures.} \textbf{a}, Non-interacting band structure with labeled $C=2$ flat band. \textbf{b}, Hartree-Fock band structure along the $k_x = 0$ line for the $\nu = 7/2$ SBCI with no remote bands. Because there is no spin polarization in valley $K_-$ the bands are two fold degenerate in that valley. The $C=2$ spin up valley $K_+$ BM band splits into the two yellow $C=1$ bands after translation breaking. \textbf{c}, Hartree-Fock band structure with five remote bands (two above $C=2$ and three below $C=2$). \textbf{d, e,} Berry curvature of the conduction and valence $C=1$ bands in Hartree-Fock with five remote bands kept. \textbf{f}, Average of \textbf{d} and \textbf{e}; the average Berry curvature has a diminished peak at the $K$ point relative to the average folded non-interacting Berry curvature shown in \textbf{g} likely due to mixing with the $C=-1$ remote band. Keeping no remote bands the two averages are necessarily the same.
    }
    \label{fig:bandsberry}
\end{figure}

\begin{figure}[ht!]
    \centering
    \includegraphics[width=\textwidth]{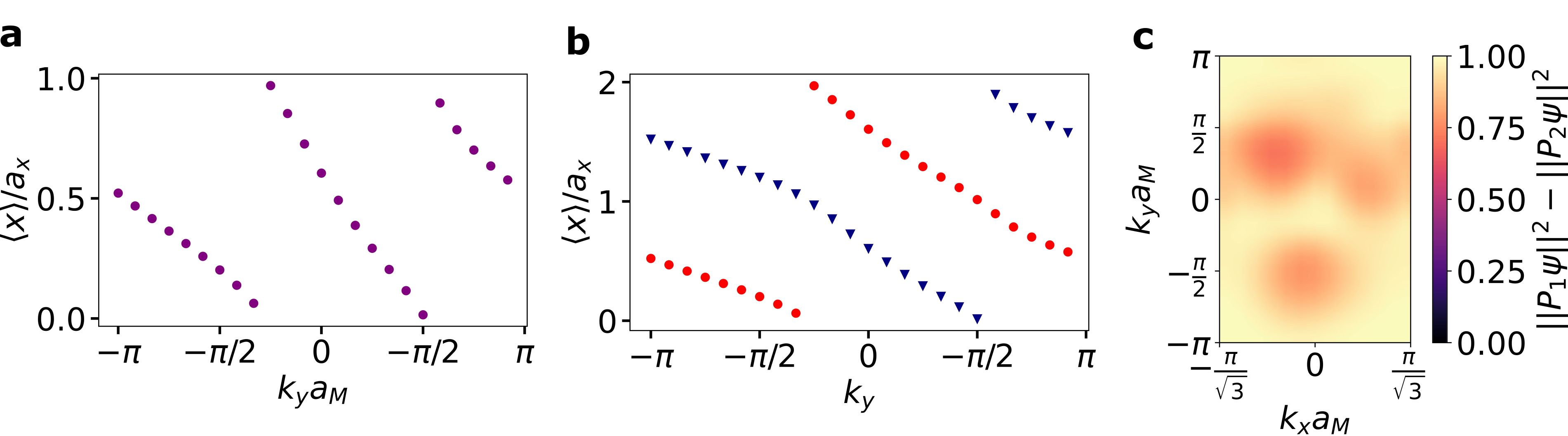}
    \caption{
    \textbf{Wannier winding and polarization.} \textbf{a}, Wannier polarization for the non-interacting $C=2$ band. The polarization winds twice as $k_ya_M$ varies from $-\pi$ to $\pi$ as expected. Here $a_x = \frac{\sqrt{3}}{2} a_M$ is the rectangular unit cell width in the $x$ direction. \textbf{b}, After doubling the unit cell, the Wannier functions split into two branches that each wind once and therefore describe $C=1$ bands. \textbf{c}, Polarization of the Hartree-Fock conduction band in the non-interacting Chern band basis of \textbf{b}.
    }
    \label{fig:wannier}
\end{figure}

We show more details about the Hartree-Fock ground state at $D= 0.4$ V/nm in Figures \ref{fig:bandsberry} and \ref{fig:wannier}. In figure \ref{fig:bandsberry}a we plot the non-interacting band structure. The $C = 2$ flat band is our band of interest; note that it comes close to touching the $C=-1$ band below it at the $K^M_+$ point. Hartree-Fock band structures are shown in Figure \ref{fig:bandsberry}b,c with and without remote bands included respectively; The $C = 2$ band in graphene valley $K_+$ with spin up splits into the two yellow bands shown after translation breaking, one of which is occupied. The other six bands, with spin down and/or graphene valley $K_-$, are all occupied giving a total filling of $\nu = 7/2$ (note that the bands in valley $K_-$ are spin degenerate since spin rotations in that valley remain unbroken). 

Adding remote bands does not change the state dramatically, although the Berry curvature at the $K^M_+$ point diminishes likely due to mixing with the $C=-1$ band below due to the near gap closing point. We may see this by comparing the average of the Berry curvature of the two $C  = +1$ Hartree-Fock bands with the average of the folded Berry curvature of the non interacting $C=2$ bands. While these averages are and should be identical when no remote bands are included, the Berry curvature at the $K^M_+$ point is diminished relative to the non-interacting case when remote bands are included.

One way to view the $C=2$ band splitting into two $C=1$ bands is through hybrid Wannier functions. While there are no fully localized Wannier functions for Chern bands, we may localize the states along the $x$ direction while keeping the state delocalized along $y$ and labeled by $k_y$ \cite{marzari_maximally_2012, qi_generic_2011}. We obtained maximally localized hybrid wannier functions and their polarization for $C=2$ band following the algorithm in \cite{marzari_maximally_1997}. In this basis the Chern number manifests itself through the winding of the polarization: as $k_y$ winds around the Brillouin zone the position of the Wannier function shifts by $-C$ unit cells in the $x$ direction as shown in Figure \ref{fig:wannier}a. Furthermore, when we double the unit cell we obtain two branches of Wannier functions that each wind once, see Figure \ref{fig:wannier}b, and therefore describe two $C=1$ bands \cite{barkeshli_topological_2012}. The ``forgotten" translation symmetry exchanges these two bands. We find that the Hartee Fock SBCI state essentially picks one of these Chern bands to fill and leaves the other unfilled; the polarization is shown in Figure \ref{fig:wannier}c and it is almost maximal everywhere in the Brillouin zone. It seems to look very similar to the plot of the translation breaking order parameter in Figure \ref{fig:theory}c as well.

\end{document}